\documentclass[journal=largetwo,manuscript=article-type,year=2024,volume=37]{cup-journal}
\usepackage{amsmath}
\usepackage{amssymb}
\usepackage{graphicx}
\usepackage{url}
\usepackage[nopatch]{microtype}
\usepackage{booktabs}
\usepackage{aas-macros}

\usepackage{hyperref} 
\hypersetup{colorlinks,citecolor=blue,linkcolor=blue,urlcolor=blue}

\title[No radio detections of passive spiral galaxies]{No detection of radio continuum from low-redshift passive spiral galaxies} 

\author{Hei Yung Chan}
\affiliation{School of Physics \& Astronomy, Monash University, Clayton, Victoria 3800, Australia}
\email[H. Y. Chan]{hcha0138@student.monash.edu}

\author{Michael J. I. Brown}
\affiliation{School of Physics \& Astronomy, Monash University, Clayton, Victoria 3800, Australia}
\begin{document}

\begin{abstract}
Radio-emitting active galactic nuclei (AGNs) are common in elliptical galaxies and AGN feedback is one of the possible mechanisms for regulating star formation in massive galaxies. It is unclear if all passive galaxy populations host radio AGNs and if AGN feedback is a plausible mechanism for truncating or regulating star formation in these galaxies. To determine if radio AGNs are common in passive spiral galaxies, we have measured the radio emission of 38 low-redshift passive spiral galaxies using RACS-low at 887.5 MHz and VLASS at 3 GHz. We selected a subset of 2MRS galaxies with negligible WISE 12 $\mu$m emission from warm dust, and spiral morphologies from HyperLeda, RC3, 2MRS and manual inspection. In contrast to comparable early-type galaxies, our sample has no significant radio detections, with radio flux densities below 1 mJy, implying that radio AGNs are rare or non-existent in passive spirals. Using the combined radio images and assuming radio luminosity is proportional to $K$-band luminosity, we find ${\rm log}~L_\nu \lesssim 9.01-0.4~M_K$. This falls below the radio luminosities of passive elliptical galaxies, implying radio luminosity in passive galaxies is correlated with host galaxy morphology and kinematics. 
 
\end{abstract}

\section{INTRODUCTION}
\label{sec:intro}

%% PARA: Radio-loud AGNs are common in massive early-types
Powerful radio-emitting AGNs are usually found in massive elliptical galaxies such as M87 \citep[e.g.][]{bolton1949} and the median radio luminosity of AGNs hosted by early-type galaxies increases with host galaxy mass \citep[e.g.][]{sadler1989, best2005, brown2011}. Radio AGNs hosted by lenticular galaxies \citep[e.g.][]{veron2001} and spiral galaxies \citep[e.g.][]{kav2015, singh2015} are rare or less luminous in radio, indicating a trend in radio luminosity with morphology and kinematics \citep[e.g.][]{hummel1983, bender1987}.

%% PARA: AGN Feedback 
Radio-mode AGN feedback, including strong winds, radio jets and radiation, could potentially regulate or truncate star formation by limiting gas supply and cooling, and this may occur with or without morphological change \citep[e.g.][]{bower2006, croton2006}. Radio lobes and radio bubbles have been observed to interact with their host galaxies, with Perseus A being the best-known example \citep[e.g.][]{boehringer1993,fabian2006, fabian2017}. While there is strong evidence for AGN feedback in individual galaxies, these powerful radio sources are often hosted by galaxies with star formation and it is not clear whether AGN feedback is a common mechanism for regulating star formation. 

%% PARA: kinematics
Host kinematics is correlated with the radio luminosity of radio AGNs. Recent work on the integral field spectroscopic data of radio galaxies by \citet{zheng2023} discovered that early-type galaxies with lower angular momenta at fixed stellar mass host stronger radio AGN sources. \citet{brown2024} also found that the slow-rotator early-type galaxies have higher median radio luminosities ($L_{1.4 \rm{~GHz}} \sim 10^{22} \mathrm{~W/Hz}$) than the fast-rotator early-type galaxies ($L_{1.4 \rm{~GHz}} \sim 10^{20} \mathrm{~W/Hz}$). Given the correlation between kinematics and radio luminosity in passive elliptical galaxies, does this trend continue to include passive spiral galaxies?

% PARA: Passive spiral
Passive spiral galaxies are a small but significant subpopulation of passive galaxies, with truncated star formation, high rotation velocity and low dispersion,   which is inferred from the velocity maps of star-forming and passive spiral galaxies \citep[e.g.][]{pak2019}. The quenching of their star formation may result from a variety of mechanisms \citep[e.g][]{becky2017}, as they occur in a variety of environments and have a range of morphological features \citep{amelia2018}. When passive spiral galaxies do show line emission, it is often consistent with LINER-like spectra \citep{amelia2016}, which could result from low-luminosity AGNs. If passive spiral galaxies frequently host AGNs, AGN feedback could explain how star formation was turned off in passive spiral galaxies without morphological change. However, if the relationship between kinematics and AGN radio luminosity extends to spiral galaxies, passive spiral galaxies will have a lower radio luminosity than passive elliptical galaxies.  

% PARA: radio data
The increasing depth of wide-field radio continuum imaging has enabled improved examinations of the radio emission from low-luminosity AGNs in nearby galaxies. While early wide-field radio continuum surveys such as the Third Cambridge survey \citep[3C; depth of $\sim$ 10 Jy at $\sim$ 200 MHz;][]{3C} detected the brightest sources, subsequent surveys have improved in both angular resolution and depth, and at the turn of the century the NRAO VLA Sky Survey \citep[NVSS;][]{NVSS} achieved an RMS depth of 0.45~mJy at 1.4~GHz. The latest wide-field continuum surveys have further improved the flux density and angular resolution limits compared to earlier works, and these surveys include the Rapid ASKAP Continuum Survey \citep[RACS;][]{RACS1, RACS2, RACS4} Band 1 (RACS-low) at 887.5~MHz, VLA Sky Survey \citep[VLASS;][]{VLASS, VLASS2} at 2-4~GHz and LOFAR Two-metre Sky Survey Data Release 2 \citep[LoTSS-DR2;][]{lotss1, lotss2, lotss3} at 120-168~MHz. VLASS has a sky coverage above declination $-40^\circ$ with an RMS depth of $\sim$140 $\mu$Jy/beam and a synthesised beam of $2^{\prime\prime}$, LoTSS-DR2 covers 27\% of the northern hemisphere with an RMS depth of 70-100 $\mu$Jy/beam and a synthesised beam of $6^{\prime\prime}$, and RACS-low covers the sky between declinations $-80^{\circ}$ and $+30^{\circ}$ with an RMS depth of  $\sim$250 $\mu$Jy/beam and a synthesised beam of $25^{\prime\prime}$. Combining deeper radio continuum imaging with samples of local galaxies enables the detection of the weakest radio sources ($\lesssim 10^{20} {\rm ~W/Hz}$), enabling radio detections of entire galaxy populations  \citep[e.g.][]{brown2024}.

% PARA: structure 
Motivated by this, we aim to determine whether passive spiral galaxies host AGN-powered radio sources using RACS-low and VLASS. The structure of this paper is as follows. Section~\ref{sec:data} covers the data extraction, sample selection by morphology and mid-infrared photometric colour cut, and radio flux density measurements from RACS-low and VLASS. Section~\ref{sec:results} provides our measurements of flux densities and luminosities for individual passive spiral galaxies and combined images of the passive spiral population. We discuss the possible relationship between radio luminosity and kinematics of passive spiral galaxies in Section~\ref{sec:discussion}, followed by the conclusion in Section~\ref{sec:conclusion}. All the catalogues, image extraction and measurements in this work are implemented using astropy \citep{astropy:2013, astropy:2018, astropy:2022}, astroquery \citep{astroquery} and Jupyter notebooks that build on code provided by Space Telescope Science Institute (STScI), CSIRO ASKAP Science Data Archive (CASDA), Canadian Astronomy Data Centre (CADC) and \citet{brown2024}. We use Vega magnitudes, adopt a \citet{imf} initial mass function (IMF), and a flat $\Lambda$ cold dark matter ($\Lambda$CDM) cosmology, with $\Omega_M=0.3$, $\Omega_\Lambda=0.7$, $\Omega_k=0$ and $H_0 = 70~\rm{km~s^{-1}~Mpc^{-1}}$. 
  
\section{Data and measurements}
\label{sec:data}

\subsection{Passive Spiral Sample}

%% PARA: 2MRS as base catalogue
To identify nearby galaxies, we used the Two Micron All Sky Redshift Survey \citep[2MRS;][]{2MRS}, which is a $K_s \leq$ 11.75 mag galaxy catalogue with 91\% sky coverage and 97.6\% redshift completeness. Using 2MRS as the base catalogue, we selected a bright subset of 5164 galaxies with $K_s$ < 10, which enabled us to measure or place strong limits on their radio luminosities with RACS-low and VLASS.

%% PARA: Morphology reference and Manually cleaned sample 
To select spiral galaxies, we used HyperLeda \citep{PGC} as the primary morphology reference followed by the older Third Reference Catalogue of Bright Galaxies \citep[RC3;][]{RC3} and then 2MRS. (While we initially used 2MRS as our primary morphology reference, we found it had a high contamination rate.) We selected spiral galaxies using the alphabetical and T type numerical morphological classifications, with -6 to -1, 0 and 1 to 9 corresponding to elliptical, lenticular and spiral galaxies respectively. Galaxies were chosen if they were S type in HyperLeda or 0 < T~type < 10 in RC3 when a HyperLeda morphology was unavailable as we want to include marginal cases that may be passive spirals. If both HyperLeda and RC3 morphologies were unavailable, galaxies with 0 < T~type < 10 in 2MRS were selected.

%% PARA: Colour Cut
We selected passive galaxies with mid-infrared colour cuts of $W2 - W3 \leq 1$ and $W1 - W2 \leq 0.5$ using photometry from the Wide-field Infrared Survey Explorer \citep[WISE;][]{WISE} ALLWISE Source Catalogue \citep[][]{ALLWISE}. Photometry in the $W1$, $W2$ and $W3$ bands was measured for each galaxy with matched elliptical apertures with semi-major axes set to 1.1 times the 2MASS Extended Source Catalogue $K_S$ isophotal semi-major axis. We did not use the WISE $W4$ band in our selection criteria as passive galaxies and galaxies with weak star formation are often too faint to be precisely measured in this band. 
The median semi-major axis of the elliptical aperture for our sample is $35^{\prime\prime}$, which at our median distance of $70~{\rm Mpc}$ corresponds to a projected radius of $12~{\rm kpc}$. While other catalogues of WISE photometry are available \citep[e.g.][]{cluver2025}, they can use smaller apertures for WISE $W3$ band which may result in aperture bias and passive colours for star-forming galaxies.

%% PARA: Colour Cut justification
Our colour cuts select passive galaxies without infrared emission from warm dust and with infrared spectra approximated by the Rayleigh-Jeans tail of stellar population spectra \citep[e.g.][]{amelia2016}. We have used simple colour cuts to select passive galaxies as they are effective, transparent and easily reproducible, whereas the alternative of specific star formation rates derived from SED fitting requires homogeneous optical and infrared photometry that is not readily available for the whole sky.
Our specific colour cuts are comparable to but stricter than those of \citet[][$W2 - W3 <1.5 {\rm~ in Vega}$]{cluver2017} which corresponds to ${\rm sSFR} \lesssim 10^{-11} \rm{~yr^{-1}}$ and \citet[][$W2 - W3 \leq2$ and $W1 - W2 \leq 0.8 {\rm~ in Vega}$]{pak2021} which correspond to ${\rm sSFR} < 10^{-10.4} \rm{~yr^{-1}}$ \citep{parkash2019}, as we aimed to reduce contamination by low-SFR spiral galaxies. 

%% PARA: galactic plane criteria
As passive spiral galaxies are a small subpopulation of all galaxies, we initially had relatively high contamination from misclassified lenticular and elliptical galaxies and contamination from spiral galaxies with star formation. We found that many galaxies close to the Galactic Plane ($|b| < 10^\circ$) and close to the Galactic Bulge ($|b| < 15^\circ$, $l < 60^\circ$ or $l > 300^\circ$) were contaminants (probably due to Galactic dust extinction in the optical) so we excluded these from the sample. The automated morphology, colour and Galactic Plane criteria result in an initial sample of 92 passive spiral galaxies, but this sample contains contaminants.

\begin{figure*}[tb!]
\centering
\includegraphics[width=0.9\textwidth]{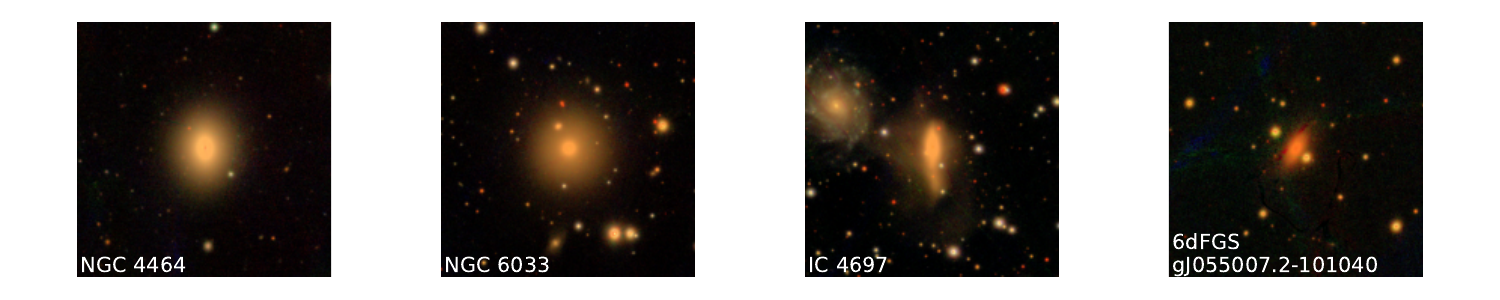}
\caption{Pan-STARRS1 $3^{\prime} \times 3^{\prime}$ \textit{grz}-images of examples of contaminants rejected by automated criteria. These galaxies have inconsistent morphologies in different catalogues or high Galactic dust extinction. NGC~4464 is a Sa in HyperLeda but a lenticular galaxy in 2MRS. NGC~6033 is a Sbc in HyperLeda but an elliptical galaxy in 2MRS. IC~4697 is peculiar in 2MRS. 6dFGS~gJ055007.2-101040 has $E(B-V) > 0.2$ and its optical image may lack morphological detail due to foreground dust extinction.}
\label{fig:contam_auto}
\end{figure*}

\begin{figure*}[tb!]
\centering
\includegraphics[width=0.9\textwidth]{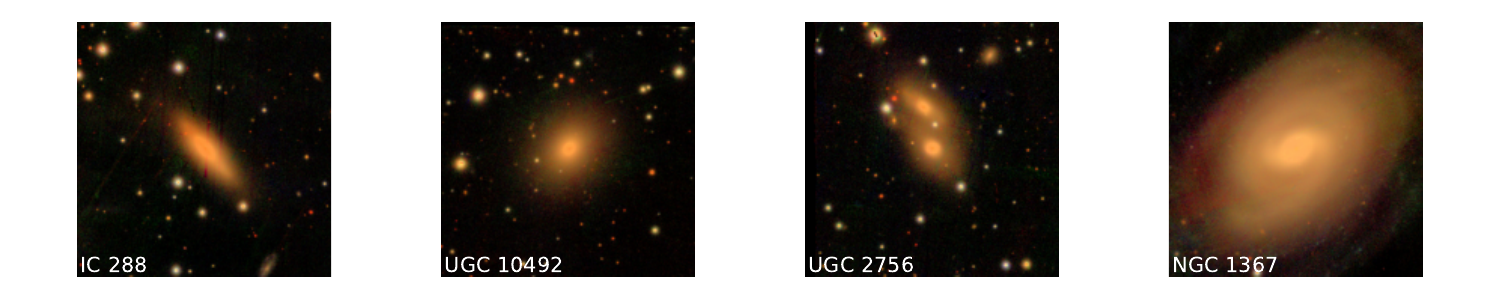}
\caption{Pan-STARRS1 $3^{\prime} \times 3^{\prime}$ \textit{grz}-images of examples of contaminants rejected by manual inspection. The contaminants are mostly misclassified elliptical, lenticular and peculiar galaxies, or spiral galaxies with low star formation. IC~288 is an edge-on lenticular galaxy with no apparent spiral arms. UGC~10492 is an elliptical galaxy with no apparent disk. UGC~2756 is a merger. NGC~1367 is a star-forming spiral galaxy that shows blue arcs at the outskirts from the Legacy Surveys. Caution that some of the images are relatively shallow and some morphological features or star formation are not apparent in Pan-STARRS1 or SkyMapper.}
\label{fig:contam_manual}
\end{figure*}

%% PARA: automated criteria to remove contaminants
To systematically identify non-spiral galaxies and remove them from our initial sample, we searched for inconsistencies between morphology catalogues and for galaxies with high Galactic dust reddening. We excluded galaxies if they had morphologies or flags in the relevant catalogues that indicated they could be peculiar, mergers or early-type galaxies. For example, NGC~6033 was classified as a spiral in HyperLeda but we excluded it from the sample because it was classified as an elliptical in 2MRS. We removed 9 galaxies with E(B-V) > 0.2 \citep{sandf2011} such as 6dFGS~gJ055007.2-101040 to avoid galaxies with potentially unreliable morphologies due to Galactic dust, and 21 galaxies with inconsistent morphologies, with example optical images in Figure~\ref{fig:contam_auto}.

%% PARA: visual inspection to remove contaminants
We visually inspected all galaxies that were selected with our morphology, colour cuts and Galactic Plane criteria, using optical images from the Panoramic Survey Telescope and Rapid Response System Telescope \#1 \citep[Pan-STARRS1;][]{PS1a, PS1b}, the Sloan Digital Sky Survey \citep[SDSS;][]{SDSS, SDSS12} or SkyMapper Southern Sky Survey \citep{SSS2}. We reject galaxies where spiral arms or disks are not clearly evident (e.g. IC~288 and UGC~10492), or where merger features (e.g. shells) have potentially been mistaken for spiral arms (e.g. UGC~2756), with example optical images in Figure~\ref{fig:contam_manual}. Apart from optical images, we looked up the NASA/IPAC Extragalactic Database \citep[NED;][]{NED} and the Legacy Surveys\footnote{\url{https://www.legacysurvey.org/viewer}} for more evidence such as multiband images \citep[e.g. GALEX FUV and NUV;][]{GALEX} and classifications from other sources to confirm the morphology and absence of star formation. Galaxies with low star formation such as NGC~1367 show blue arcs at the outskirts in Legacy Surveys and are rejected.

\begin{figure*}[tb]
     %\centering
     \includegraphics[width=0.499\textwidth]{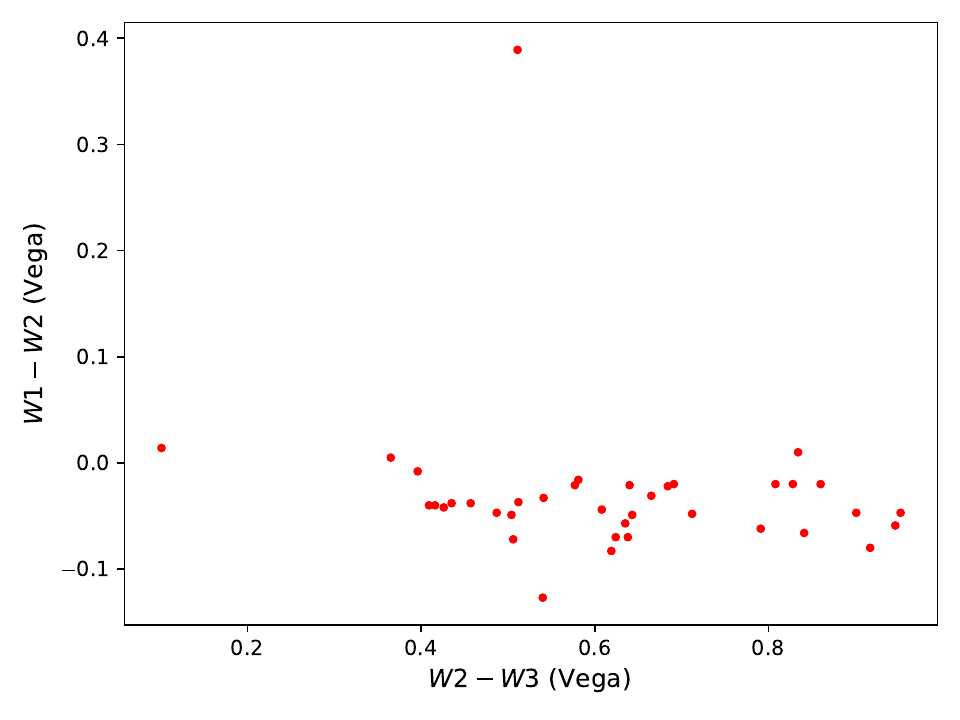}\includegraphics[width=0.499\textwidth]{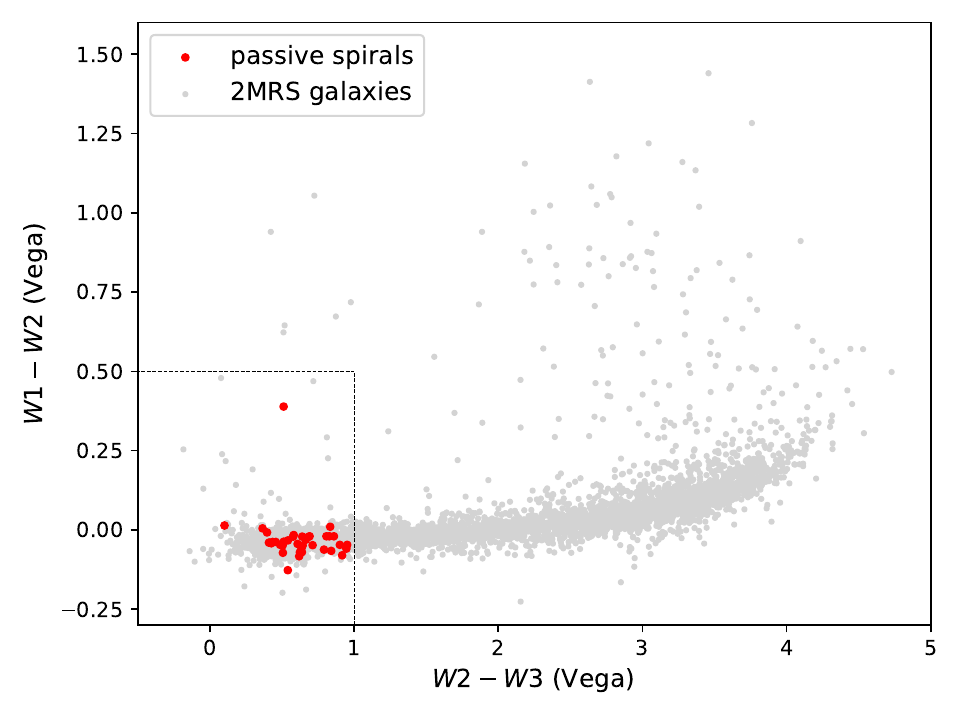}
     \caption[Colour-colour diagrams of our sample]{Colour-colour plots of our sample and 2MRS galaxies. Left: $W1-W2$ against $W2-W3$ for our passive spiral galaxies, which have been selected with $W2 - W3 \leq 1$ and $W1 - W2 \leq 0.5$. The outlier with high $W1 - W2$ is UGC 3855 with its photometry affected slightly by a neighbouring star. Right: the WISE colours of passive spirals compared to all 2MRS galaxies (light grey). Passive spiral galaxies and contaminants are located at the left, where SEDs approximate Rayleigh-Jeans spectra.}
    \label{fig:CCplot}
\end{figure*}

\begin{figure*}
\centering
\includegraphics[width=0.9\textwidth]{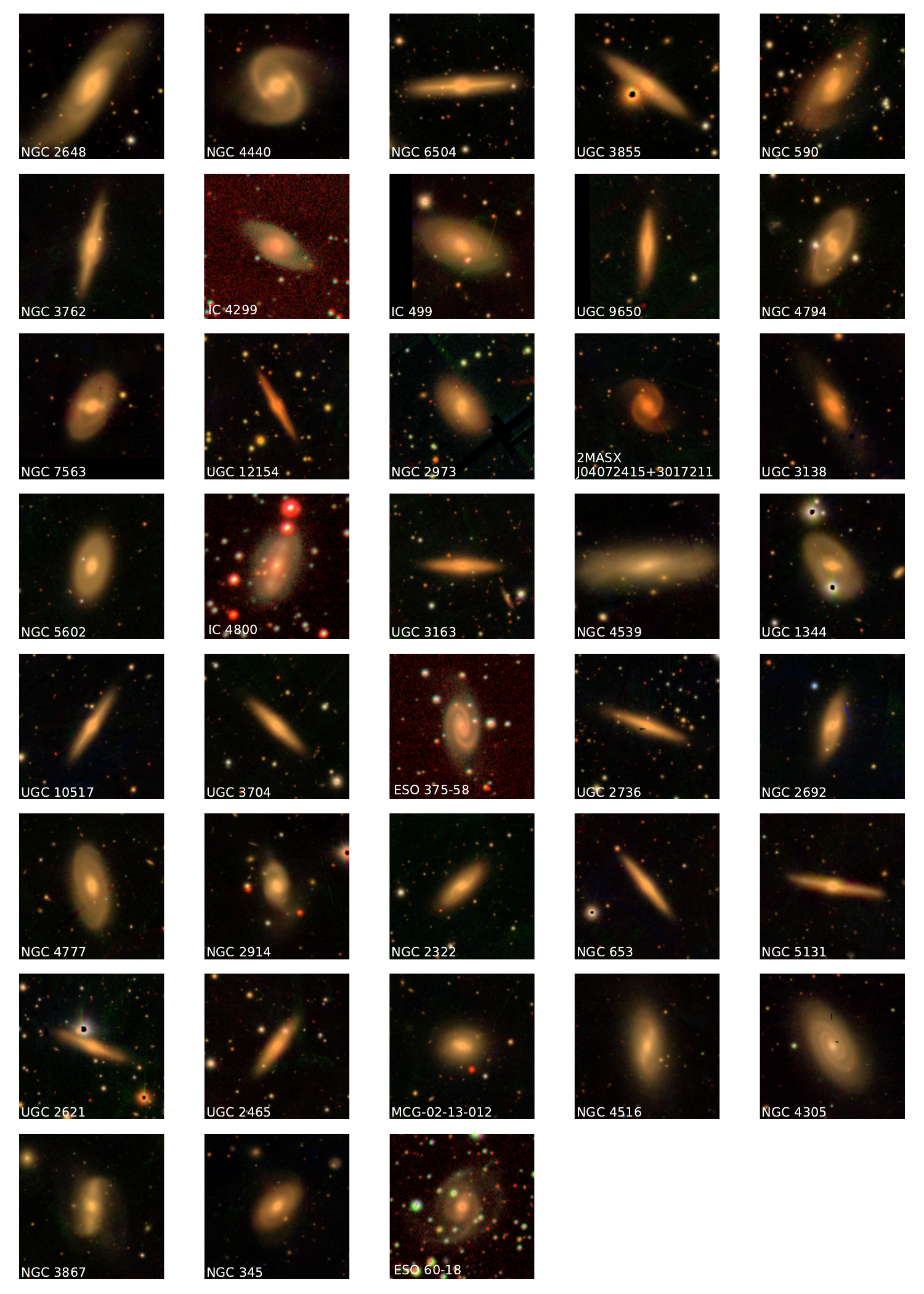}
\caption{Pan-STARRS1 and SkyMapper $3^{\prime} \times 3^{\prime}$ \textit{grz}-images of our sample of 38 passive spiral galaxies. NGC~4440 (the second one in the first row) is a face-on SBa and UGC~12154 (the second one in the third row) is an edge-on Sb in HyperLeda.}
\label{fig:sample}
\end{figure*}

%% PARA: final sample
After the manual removal of 6 galaxies with star formation and 18 galaxies with morphologies other than spirals, we have a final sample of 38 passive spiral galaxies with $8.65<K_S<9.99$. Figure~\ref{fig:CCplot} illustrates that the WISE colours of the sample are consistent with those of passive galaxies. Figure~\ref{fig:sample} provides Pan-STARRS1 and SkyMapper colour images of all 38 galaxies in the sample.

\subsection{Radio Data}

%% PARA: ASKAP and VLASS Cutouts 
To measure the radio continuum flux densities of the passive spiral galaxies, we extracted $10^{\prime} \times 10^{\prime}$ RACS-low and $5^{\prime} \times 5^{\prime}$ VLASS image cutouts with the CASDA and CADC astroquery packages respectively. To quantify the noise in the images, we have used the median absolute deviation (MAD), which we have then divided by 0.6745 to provide $\sigma$ so that it is equivalent to the root mean square (RMS) for a Gaussian noise distribution. We found that the median MAD values for the RACS-low and VLASS images were 250 $\mu$Jy and 62 $\mu$Jy respectively, which correspond to RMS values of 370 $\mu$Jy and 91 $\mu$Jy. All 38 passive spiral galaxies have at least 1 radio continuum image, with 18 galaxies having RACS-low images and 36 galaxies having VLASS images. 

%% PARA: radio measurements
We measured the radio flux density, $F_\nu$, using the pixel in the radio image corresponding to the position of the centre of each galaxy. Combined images are available for galaxies with RACS-low images. For galaxies with multiple VLASS images, median-combined images with overlapping regions of at least $300\times300$ pixels or otherwise the image with the lowest noise was used. Key measurements for individual galaxies are provided in Table~\ref{tab:RACS}, including RACS-low and VLASS flux densities per beam, RACS-low and VLASS noise, $K$-band absolute magnitude and luminosity distance. We caution that our radio flux density measurements are generally not significant and should not be confused with detections, but the distribution of measurements is useful for understanding the radio properties of the passive spiral population.

\begin{table*}[tbh!]
\begin{threeparttable}
\caption{The RACS-low sample of nearby passive spiral galaxies (full table is available online).}
\label{tab:RACS}
\scriptsize
\centering
\begin{tabular}{lccccccccc}
\toprule
\headrow  Name & R.A.     & Decl.   &  $F_{\rm RACS-low}$  & $\sigma_{\rm RACS-low}$   & $F_{\rm VLASS}$   & $\sigma_{\rm VLASS}$ & $M_K$ & $D_L$ \\ 
\headrow      & (J2000)  & (J2000)   &  (mJy) & (mJy) & (mJy)  & (mJy) & (mag) & (Mpc)\\ 
\midrule
NGC 2648 & 130.6659 & 14.2855 & 0.63 & 0.64 & 0.21 & 0.10 & -23.71 & 29.59  \\
NGC 4440 & 186.9732 & 12.2932 & 7.12 & 4.80 & -0.16 & 0.10 & -21.18 & 10.36  \\
NGC 6504 & 269.0238 & 33.2084 & - & - & -0.08 & 0.08 & -25.00 & 69.24  \\
UGC 3855 & 112.0542 & 58.5067 & - & - & 0.02 & 0.08 & -24.17 & 45.61  \\
NGC 590 & 23.4205 & 44.9287 & - & - & 0.08 & 0.09 & -25.08 & 71.70  \\
NGC 3762 & 174.3491 & 61.7594 & - & - & -0.18 & 0.08 & -24.14 & 49.91  \\
IC 4299 & 204.1982 & -34.0659 & -1.43 & 1.10 & 0.10 & 0.15 & -24.50 & 58.75  \\
IC 499 & 131.3212 & 85.7400 & - & - & -0.08 & 0.09 & -22.80 & 27.02  \\
UGC 9650 & 223.4563 & 83.5905 & - & - & -0.01 & 0.09 & -24.33 & 56.29  \\
NGC 4794 & 193.7937 & -12.6085 & 0.67 & 0.72 & -0.10 & 0.13 & -24.43 & 57.80  \\
NGC 7563 & 348.9831 & 13.1962 & -0.73 & 0.87 & -0.15 & 0.09 & -24.53 & 62.06  \\
UGC 12154 & 340.1494 & 72.8639 & - & - & 0.17 & 0.09 & -24.36 & 60.21  \\
NGC 2973 & 144.4984 & -30.1487 & 0.13 & 0.25 & -0.05 & 0.08 & -24.51 & 63.41  \\
2MASX J04072415+3017211 & 61.8506 & 30.2892 & - & - & -0.06 & 0.09 & -25.03 & 80.59  \\
UGC 3138 & 70.7150 & 18.5546 & -0.30 & 0.34 & -0.06 & 0.09 & -24.69 & 69.73  \\
NGC 5602 & 215.5784 & 50.5014 & - & - & 0.20 & 0.08 & -22.94 & 32.04  \\
IC 4800 & 284.6813 & -63.1392 & 0.17 & 0.37 & - & - & -24.52 & 65.00  \\
UGC 3163 & 72.3895 & 69.4750 & - & - & 0.00 & 0.09 & -24.50 & 69.13  \\
NGC 4539 & 188.6448 & 18.2026 & 1.25 & 0.37 & -0.09 & 0.09 & -22.07 & 20.03  \\
UGC 1344 & 28.1446 & 36.5009 & - & - & 0.19 & 0.08 & -24.22 & 60.21  \\
UGC 10517 & 250.2877 & 61.3263 & - & - & 0.28 & 0.09 & -24.70 & 81.08  \\
UGC 3704 & 107.6283 & 61.7863 & - & - & 0.08 & 0.09 & -23.72 & 52.11  \\
ESO 375-58 & 157.9822 & -35.4098 & 0.19 & 0.48 & 0.03 & 0.09 & -23.19 & 38.91  \\
UGC 2736 & 51.6149 & 40.5079 & - & - & 0.05 & 0.09 & -24.76 & 85.47  \\
NGC 2692 & 134.2418 & 52.0660 & - & - & -0.04 & 0.07 & -23.79 & 54.42  \\
NGC 4777 & 193.4939 & -8.7757 & 0.69 & 0.50 & 0.21 & 0.10 & -23.63 & 51.19  \\
NGC 2914 & 143.5116 & 10.1088 & -0.56 & 0.50 & 0.15 & 0.10 & -23.36 & 45.28  \\
NGC 2322 & 106.5012 & 50.5103 & - & - & -0.11 & 0.09 & -24.90 & 91.78  \\
NGC 653 & 25.6072 & 35.6383 & - & - & 0.08 & 0.08 & -24.51 & 79.21  \\
NGC 5131 & 200.9873 & 30.9880 & - & - & 0.17 & 0.12 & -25.04 & 99.52  \\
UGC 2621 & 49.1089 & 41.5304 & - & - & 0.19 & 0.33 & -24.25 & 68.64  \\
UGC 2465 & 45.1559 & 35.1691 & - & - & 0.04 & 0.13 & -24.35 & 73.49  \\
MCG-02-13-012 & 72.4279 & -10.7068 & -0.55 & 0.72 & -0.06 & 0.12 & -23.98 & 60.52  \\
NGC 4516 & 188.2814 & 14.5749 & 2.63 & 1.74 & 0.07 & 0.09 & -20.68 & 13.49  \\
NGC 4305 & 185.5150 & 12.7409 & -0.45 & 0.76 & 0.09 & 0.09 & -22.39 & 27.58  \\
NGC 3867 & 176.3735 & 19.4002 & 0.19 & 0.61 & 0.01 & 0.09 & -25.20 & 108.88  \\
NGC 345 & 15.3421 & -6.8843 & 0.08 & 0.39 & 0.07 & 0.10 & -24.42 & 75.72  \\
ESO 60-18 & 134.1689 & -67.8703 & 0.31 & 0.23 & - & - & -25.20 & 103.51  \\
\bottomrule
\end{tabular}
\end{threeparttable}
\end{table*}

\section{Results}
\label{sec:results}

\begin{figure*}[tb]
    % \centering
    \includegraphics[width=0.499\textwidth]{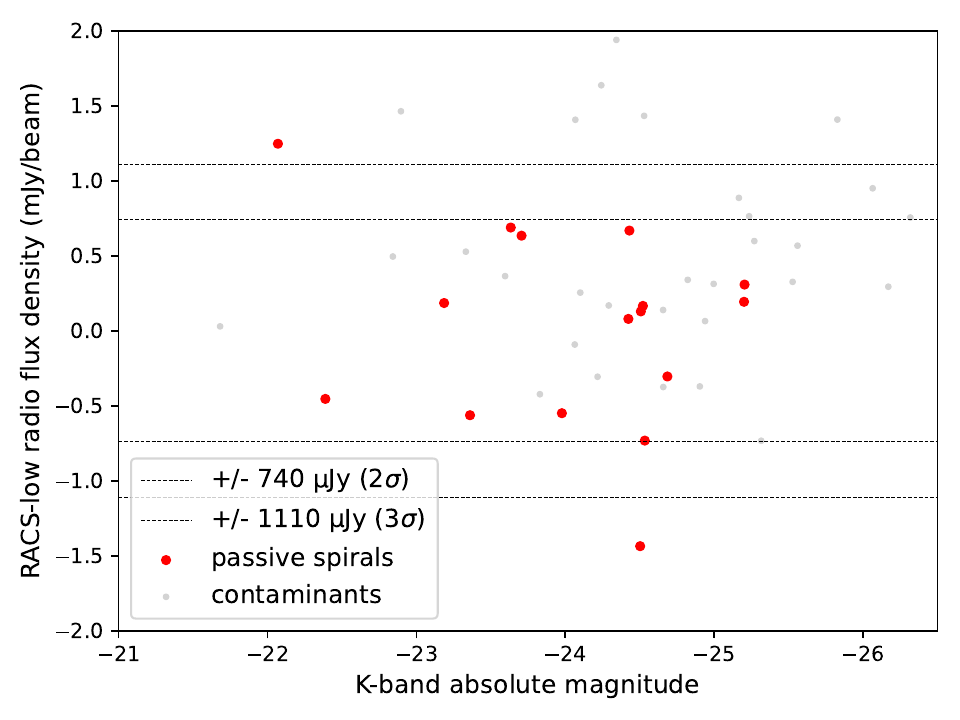}\includegraphics[width=0.499\textwidth]{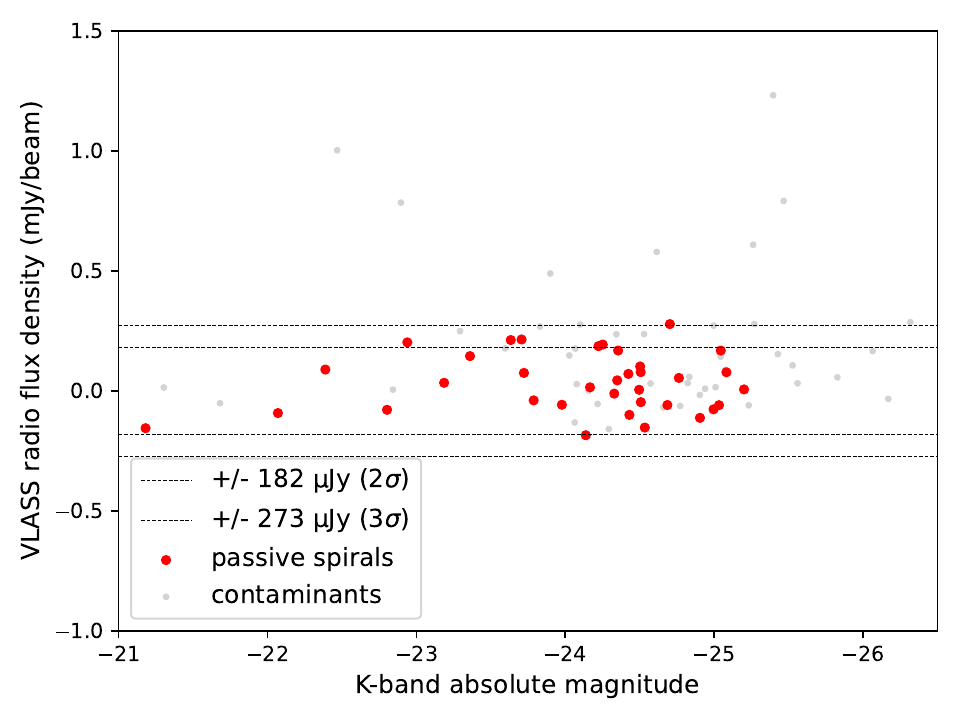}
    \caption[Radio flux density versus $K$-band absolute magnitude]{RACS-low (left) and VLASS (right) radio flux densities as a function of $K$-band absolute magnitude (excluding 2 galaxies with noisy RACS-low images). The radio flux densities of passive spiral galaxies (red) are generally lower than those of the star-forming and early-type contaminants rejected from the sample (light grey). The median RMS errors for the RACS-low and VLASS images are 370 $\mu$Jy and 91 $\mu$Jy respectively. After excluding images with noisy data and visual inspection of images with $2\sigma$ measurements, we conclude there are no detections of our passive spiral galaxies with RACS-low and VLASS.}
    \label{fig:fnu_K}
\end{figure*}

%% PARA: Flux Density significance of passive spirals
None of the passive spiral galaxies in our sample have significant ($>3\sigma$) detections in RACS-low and VLASS except two galaxies. NGC~4539 has a $3.34 \sigma$ detection in RACS-low and UGC~10517 has a $3.21 \sigma$ detection in VLASS but both galaxies are contaminated by neighbouring sources. We visually inspected 4 galaxies with $2-3\sigma$ measurements and their images are consistent with noise rather than radio sources.
Figure~\ref{fig:fnu_K} shows the radio continuum flux density measurements for the passive spiral galaxies. The numbers of positive and negative measurements are comparable, indicating that the measurements of passive spiral galaxies are consistent with a noise distribution centred on zero. All but 3 of the RACS-low measurements (NGC~4440, NGC~4516 and NGC~4539) are consistent with flux densities below 1~mJy, with the first two exceptions being in images with high noise levels of $>1 {\rm ~mJy}$ and the last one (NGC~4539) contaminated by a neighbouring source. All VLASS measurements are consistent with flux densities below 1~mJy. We conclude that there are no valid radio sources above $3\sigma$ in either RACS-low or VLASS for the passive spiral sample.

\begin{figure}[tb]
     \centering
     \includegraphics[width=\textwidth]{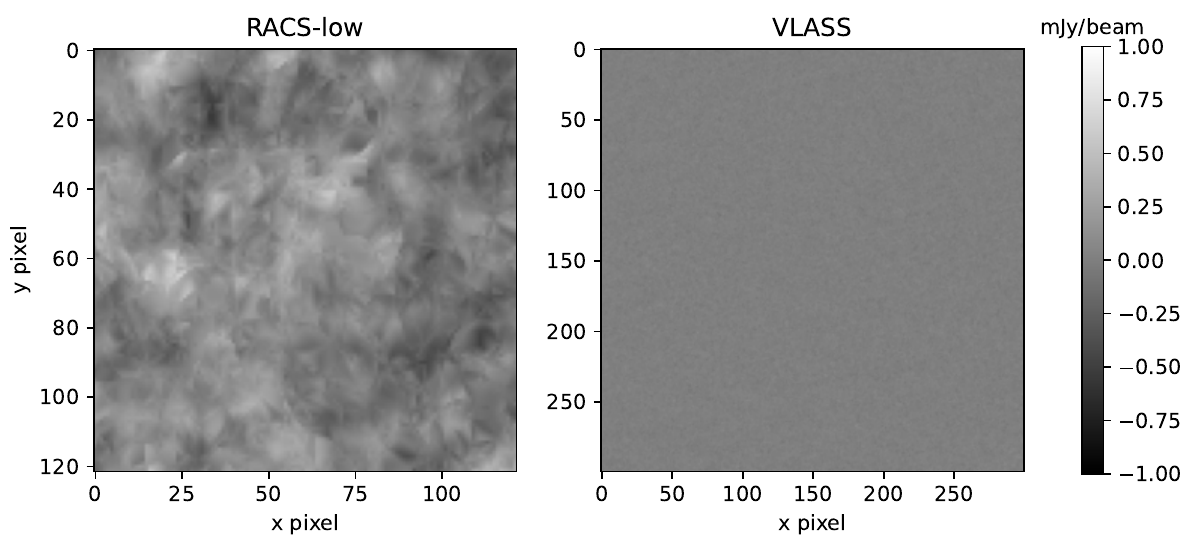}
    \caption[radio combined images]{Combined radio continuum images of passive spiral galaxies. Left: 17 RACS-low images were combined to form this image with $ F_\nu = 167 ~\mu$Jy and $\sigma = 161 ~\mu$Jy. Right: 31 VLASS images were combined to form this image with $ F_\nu = 6.31 ~\mu$Jy and $\sigma = 20.5 ~\mu$Jy. There is no significant radio emission in the RACS-low and VLASS combined images of passive spiral galaxies. } 
    \label{fig:combined_images}
\end{figure}

%% PARA: combined image
To search for weak radio emission from the passive spiral galaxy population, we median combined 17 RACS-low and 31 VLASS images (excluding 1 RACS-low image and 5 VLASS images with partial coverage of the field-of-view, or contaminated by streaks or neighbouring sources). The resulting combined radio continuum images are shown in Figure~\ref{fig:combined_images}. The RACS-low combined image has a central flux density of 167 $\mu$Jy and RMS of 161 $\mu$Jy while the VLASS one has a central flux density of 6.31 $\mu$Jy and RMS of 20.5 $\mu$Jy. Both central flux densities are lower than 2 times the relevant RMS, consistent with random noise and our passive spiral galaxies typically having RACS-low flux densities of $\lesssim 200~\mu {\rm Jy}$.

\begin{figure}[tb]
    \centering
    \includegraphics[width=\textwidth]{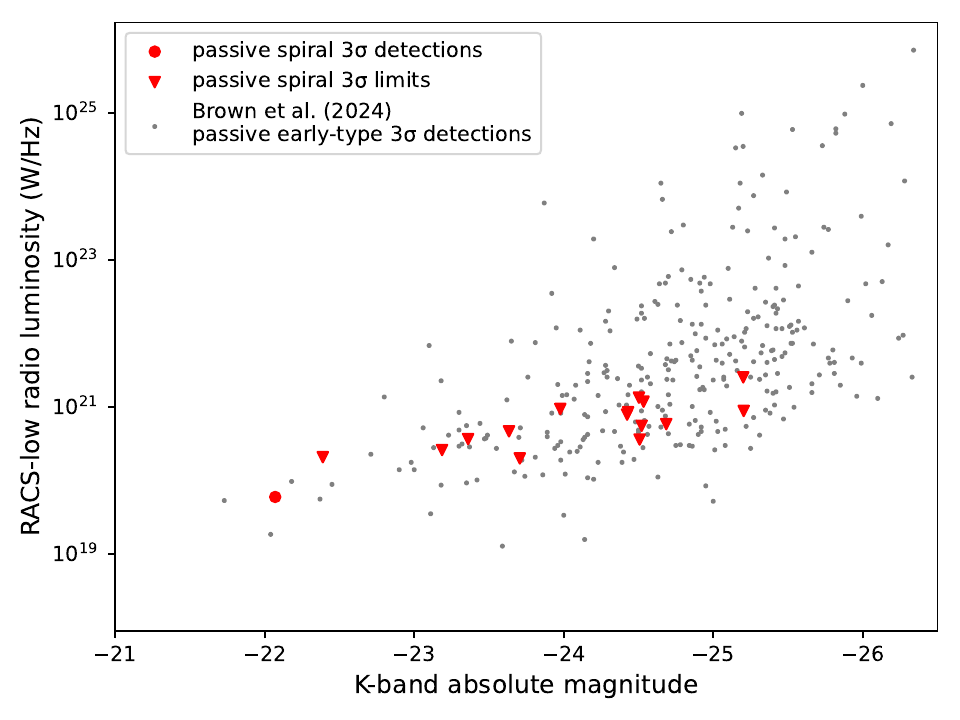}
    \caption[Radio luminosity as a function of $K$-band absolute magnitude, compared to passive elliptical galaxies]{Radio luminosity as a function of $K$-band absolute magnitude for passive spiral galaxies (red) and the early-type sample (grey) of \citet{brown2024}. Galaxies with flux densities within $3\sigma$ of zero are shown with upper limits and for clarity galaxies with very noisy images are excluded from the figure. Our sample has lower radio luminosities than most early-type galaxies, implying that passive spiral galaxies are less luminous in radio than comparable passive elliptical and lenticular galaxies. }
    \label{fig:Lnu_K_all}
\end{figure}

%% PARA: Radio luminosity compared to brown 2024
Figure~\ref{fig:Lnu_K_all} compares the radio luminosities of our passive spiral galaxies with the radio luminosities of early-type galaxies from \citet{brown2024}, which are also measured with RACS-low. When an insignificant flux density was measured for a galaxy, we determined an upper limit for the luminosity using a flux density corresponding to three times RMS. The radio luminosity upper limits of our passive spiral galaxies are comparable to or less than $10^{21} \mathrm{~W/Hz}$, which is lower than the typical radio luminosity of early-type galaxies. The early-type galaxies of \citet{brown2024} are mostly passive with radio sources powered by AGNs, and Figure~\ref{fig:Lnu_K_all} suggests such radio sources are rare, weak or absent from passive spiral galaxies.

\section{DISCUSSION}
\label{sec:discussion}

%% PARA: no evidence of radio AGN in passive spiral galaxy - luminosity VS morph
We find no evidence of radio emission from passive spiral galaxies, and Figure~\ref{fig:Lnu_K_all} shows that the radio luminosities of passive spiral galaxies are far weaker (on average) than those of elliptical and lenticular galaxies. Radio-loud AGNs are usually found in passive elliptical galaxies but rarely in star-forming spiral galaxies \citep[e.g.][]{kormendy2013}, suggesting AGN radio luminosity is correlated with morphology or anti-correlated with star formation rate. Our finding indicates the importance of the correlation between radio luminosity and morphology. 

\begin{figure}[tb]
    \centering
    \includegraphics[width=\textwidth]{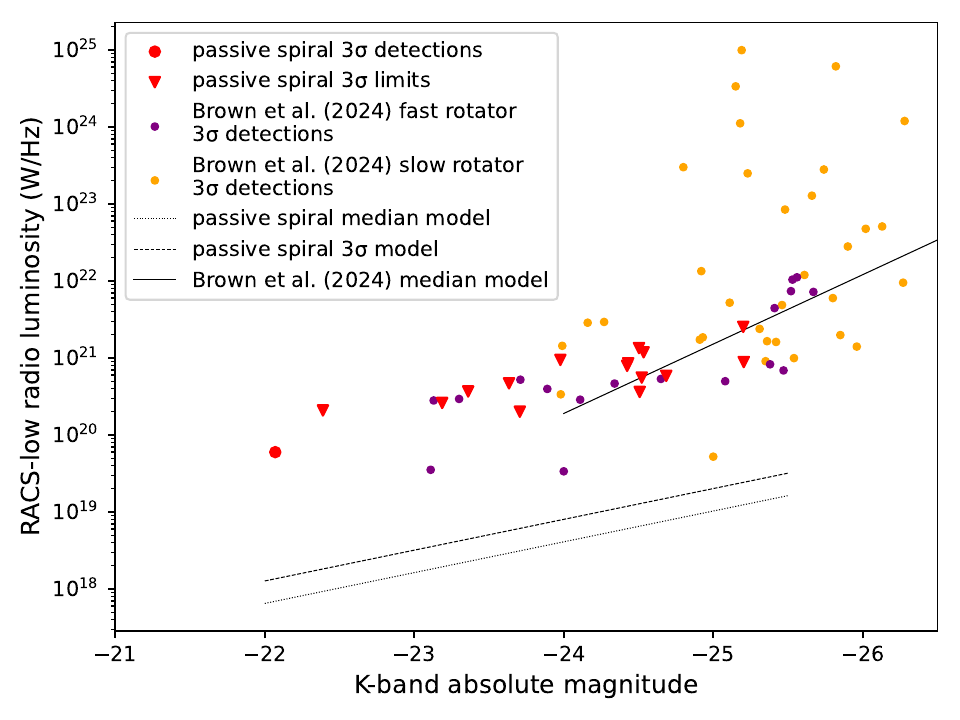}
    \caption[887.5 MHz radio luminosity versus $K$-band absolute magnitude compared to passive elliptical galaxies]{Radio luminosity as a function of $K$-band absolute magnitude for passive spiral galaxies, compared with fast-rotating (purple) and slow-rotating (yellow) early-type galaxies from \citet{brown2024}. Galaxies with a measured flux density per beam less than $3\sigma$ above the local noise are plotted as triangles using the values of the $3\sigma$ upper limits. The dotted line and dashed line correspond to the best fits of the central radio flux density and $3\sigma$ upper limit of the combined RACS-low image of passive spiral galaxies, assuming radio luminosity is proportional to $K$-band luminosity. The $3\sigma$ line is lower than the passive early-type median (solid line) and the radio luminosities for all slow rotator and fast rotator early-type galaxies.}
    \label{fig:Lnu_K_all_FS}
\end{figure}

%% PARA: luminosity model - morph trend
The combined radio images better constrain the radio luminosities of passive spiral galaxies than the individual images. To model the radio luminosities of passive spiral galaxies, we assume radio luminosity (in ${\rm W/Hz}$) is directly proportional to the $K$-band luminosity, which is equivalent to 
\begin{equation}
    {\rm log~} L_\nu = A - 0.4 M_K,
\end{equation}
where $A$ is a constant and $M_K$ is the $K$-band absolute magnitude between -22 and -25.5. The values of A using the central flux density and $3\sigma$ upper limit from the combined RACS-low image are $A_{avg} = 9.01$ (dotted line) and $A_{3\sigma} = 9.30$ (dashed line) respectively as shown in Figure~\ref{fig:Lnu_K_all_FS}. The $3\sigma$ line is below the passive early-type median line (solid line), all $3\sigma$ detections of slow-rotating (yellow) and fast-rotating (purple) early-type galaxies from \citet{brown2024}, indicating that passive spiral galaxies have lower radio luminosities than both fast- and slow-rotating passive early-types.

%% PARA: morph differences in radio powers - host galaxy kinematics
The radio luminosities of passive galaxies have a dependence on morphology, with elliptical galaxies having higher radio luminosities than lenticular galaxies (on average) and lenticular galaxies having higher radio luminosities than passive spiral galaxies (on average). There could be a direct connection between radio luminosity and morphology, or morphology could be correlated with the underlying causal relationship. For example, radio luminosity could be a function of host galaxy stellar kinematics. Slow rotator early-type galaxies typically have higher radio luminosities than fast rotator early-type galaxies \citep{zheng2023, brown2024}, and passive spiral galaxies are almost certainly fast rotators with lower velocity dispersions than elliptical and lenticular galaxies \citep[e.g.][]{pak2019}. Irrespective of the relationship between radio luminosity and morphology, the absence of radio emission in passive spiral galaxies indicates that radio-mode AGN feedback is not regulating star formation in these galaxies and that radio-loud AGNs are not necessarily common in all passive galaxy populations.

%% PARA: compare to rowlands2012
Our result of no significant radio emission from passive spiral galaxies is consistent with earlier work by \citet{rowlands2012}. They found no radio matches for their $0.05 \le z < 0.25$ passive spiral sample in 1.335~GHz FIRST images that have a median RMS depth of $\sim 100~\mu$Jy/beam. \citet{rowlands2012} median galaxy redshift is approximately an order of magnitude higher than our own, and thus their radio luminosity sensitivity is roughly two orders of magnitude poorer than our work. Overall, no significant radio emission is detected from passive spiral galaxies in RACS-low, VLASS and FIRST.

%% PARA: possibility of radio source
While we find no evidence for radio emission from passive spiral galaxies, and we conclude that passive spiral galaxies must typically have far lower radio luminosities than comparable elliptical and lenticular galaxies, we cannot rule out radio sources in this population. We have a sample of only 38 passive spiral galaxies, so ~1\% of passive spiral galaxies hosting $\sim 10^{20} \mathrm{~W/Hz}$ radio sources would be entirely consistent with our work. Such radio sources powered by AGNs are found in adjacent populations, including passive lenticular galaxies \citep[e.g.][]{veron2001} and spiral galaxies with low (but non-zero) star formation rates \citep[e.g. NGC~1367,][]{condon2019}. While we have searched for radio emission from AGNs in passive spiral galaxies, other sources could contribute to the radio emission from these galaxies, including populations of millisecond pulsars, which are predicted to produce radio luminosities on the order of $\sim 10^{21} \mathrm{~W/Hz}$ \citep[][]{sudoh2021}.

%% PARA: X-ray & nebular emission --> lack of AGN in psg
While the lack of radio emission in passive spiral galaxies indicates an absence of radio-mode AGNs, they could host radio-quiet AGNs. To see if they host radio-quiet AGNs, we searched for their X-ray counterparts. We cross-matched the passive spiral sample and (for comparison) the early-type sample from \citet{brown2024} with the main X-ray catalogue from eROSITA/eRASS1 \citep{erass} at 0.2–2.3~keV using a searching radius of $15^{\prime\prime}$. We identified only 1 X-ray source (IC~4299 with $F_{\rm 0.2-2.3~keV} = 1.23 \times 10^{-13} \rm{~erg~s^{-1}~cm^{-2}}$) in passive spiral galaxies (out of 16 galaxies in the eROSITA/eRASS1 area) and 259 matches (out of 484 galaxies in the eROSITA/ eRASS1 area) in the early-type sample, indicating that X-ray detectable AGNs are rare in passive spiral galaxies but far more common in early-type galaxies. (Although we also caution that X-ray emission can originate from sources other than AGNs in galaxies \citep[e.g.][]{white2002, fabbiano2006}.) There is no obvious evidence indicating that IC~4299 is an AGN as its WISE colours are located outside the AGN wedge and its 6dF optical spectrum \citep[][]{jones2009} does not show AGN features like emission lines or broad lines. Moreover, the rarity of 
prominent emission lines observed in passive spiral galaxies indicates weak or no AGN activity \citep{masters2010, rowlands2012, amelia2016}. Therefore, we see no evidence for radio emission from AGNs in the passive spiral population, while X-ray and nebular emission lines suggest AGNs hosted by this population are relatively rare or weak.

\section{CONCLUSION}
\label{sec:conclusion}

%% PARA: methodology

We have searched for radio emission from passive spiral galaxies to determine if they host radio sources powered by AGNs, as do many passive elliptical galaxies. To do this we selected a sample of 38 low-redshift $K_S<10$ passive spiral galaxies from 2MRS using WISE infrared colour cuts of $W2 - W3 \leq 1$, $W1 - W2 \leq 0.5$ and morphologies from HyperLeda, RC3, 2MRS, Galactic coordinates cut, $E(B-V)$ cut and manual inspection. We then measured their radio continuum flux densities using RACS-low and VLASS images. 

%% PARA: results

We found no significant radio emission from our passive spiral sample. Their flux densities are all within $3\sigma$ of zero and the exceptions are consistent with noise or contamination in the radio continuum images. The central flux densities from the stacked RACS-low and VLASS images of the passive spiral galaxy population are also consistent with noise. Assuming radio continuum luminosity is proportional to the $K$-band luminosity, the average RACS-low radio luminosity is $ L_\nu = 10^{9.01-0.4 M_k} \lesssim 10^{20} \mathrm{~W/Hz}$, which is lower than the radio luminosities of passive early-type galaxies. 

% PARA: morphological trend in radio luminosity
Passive spiral galaxies on average have lower radio luminosities than slow- and fast-rotator passive early-type galaxies. We conclude that radio emission from passive galaxies is correlated with kinematics and morphology, although we cannot identify the direct causal link. While radio-mode AGN feedback may play a role in truncating and regulating star formation in some passive galaxies, the lack of radio-emitting AGNs in passive spiral galaxies suggests that AGN feedback is not significant in these galaxies. 

\begin{acknowledgement}
This work used data from RACS, VLASS, WISE, HyperLeda, RC3, 2MRS, Pan-STARRS1, SDSS, SkyMapper, Legacy Surveys and NED. The Australian Square Kilometre Array Path\-finder is part of the Australia Telescope National Facility which is managed by CSIRO. Operation of ASKAP is funded by the Australian Government with support from the National Collaborative Research Infrastructure Strategy. ASKAP uses the resources of the Pawsey Supercomputing Centre. The establishment of ASKAP, the Murchison Radio-astronomy Observatory, and the Pawsey Supercomputing Centre are initiatives of the Australian Government, with support from the Government of Western Australia and the Science and Industry Endowment Fund. We acknowledge the Wajarri Yamatji as the traditional owners of the Murchison Radio-astronomy Observatory site. The National Radio Astronomy Observatory is a facility of the National Science Foundation operated under a cooperative agreement by Associated Universities, Inc. CIRADA is funded by a grant from the Canada Foundation for Innovation 2017 Innovation Fund (Project 35999), as well as by the Provinces of Ontario, British Columbia, Alberta, Manitoba and Quebec. This paper makes use of services or code that have been provided by STScI, CASDA and CADC. We thank an anonymous referee whose comments and suggestions significantly improved the manuscript.

\end{acknowledgement}

\paragraph{Data Availability Statement}

The code and complete Table \ref{tab:RACS} (in CSV format) is attached to the source file.

\begin{figure*}[bht!]
     \includegraphics[width=\textwidth]{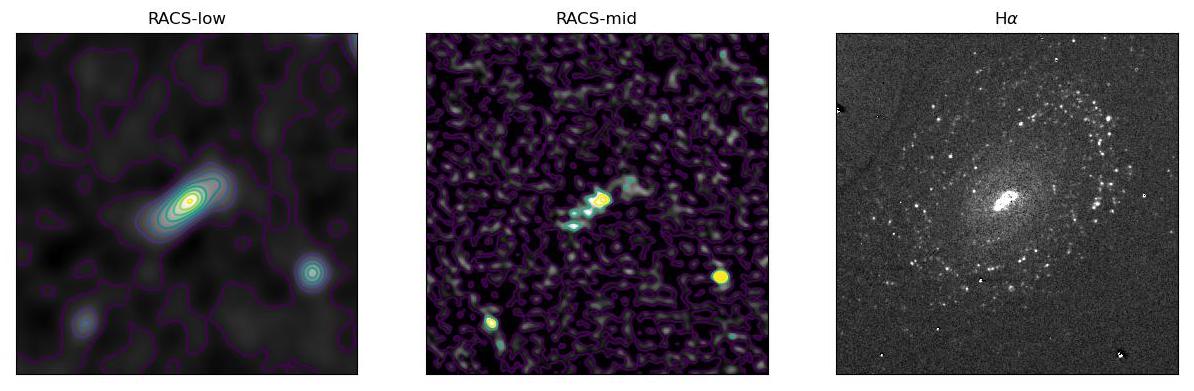}
     \caption[Radio maps and H-alpha image of NGC~1367]{Left: radio map of NGC~1367 at 887.5~MHz from RACS-low. Centre: radio map of NGC~1367 at 1367.5~MHz from RACS-mid. Right: continuum-subtracted H$\alpha$ image of NGC~1367 from \citet{hameed1999}. The position of radio emission is not consistent with optical emission, suggesting that the radio emission is powered by an AGN instead of star formation. Traces of little star formation are also seen in the spiral arms in the H$\alpha$ image. (All three images are matched in the size of $6.142^{\prime} \times 6.142^{\prime}$.)} 
     \label{fig:NGC_1371}
\end{figure*}

%\endnote in some journals will behave like \footnote; and \printendnotes will not output anything. 
\printendnotes

\printbibliography

\appendix

\section{Radio Detection of a Low-SFR Spiral}

%% PARA: NGC 1371 is an AGN
Although we do not detect radio emission from passive spiral galaxies, we do observe AGN-powered radio emission from NGC~1367 (NGC~1371), a low-SFR SBa spiral galaxy. Its RACS-low radio flux density at 887.5 MHz, measured with an elliptical aperture, is 19.7~mJy, which is very high compared to the passive spiral galaxies and the other low-SFR spiral galaxies excluded from our sample. Its RACS-low and RACS-mid (1367.5~MHz) radio images in Figure~\ref{fig:NGC_1371} show that it has strong central radio emission with double radio lobes and jets, which are consistent with the findings of \citet{omar2005} using GMRT radio continuum imaging and \citet{grundy2023} using the WALLABY survey. \citet{condon2019} also identified NGC~1367 as an AGN-powered radio source on the basis of its mid-IR to radio flux ratio, using WISE, IRAS and NVSS data. The 6dFGS optical spectrum \citep{jones2009} of the central $3.4^{\prime\prime}$ of NGC~1367 shows [N~II] in emission and H${\alpha}$ in absorption, which suggests the nuclear line emission in NGC~1367 is from a LINER or a low-luminosity AGN. NGC~1367 highlights that a subset of spiral galaxies with low SFRs have radio sources powered by AGNs.

%% PARA: NGC 1371 has low star formation
Despite its passive infrared colours, NGC~1367 shows evidence of star formation in optical images and was hence classified as a contaminant of our passive spiral sample. Traces of star formation are seen at the outskirts of the galaxy in the continuum-subtracted H${\alpha}$ (and [N~II]) image from \citet{hameed1999}, which we show in Figure~\ref{fig:NGC_1371}, and in $grz$ images from Legacy Surveys DR9. NGC~1367 has a low SFR, with a WISE $W3$ SFR of just $0.1~M_\odot \rm{~yr^{-1}}$ \citep{cluver2014} even without subtracting the stellar continuum from the NGC~1367 SED. 

%% PARA: NGC 1371 may be quenched by AGN feedback
While it is possible that radio mode AGN feedback is regulating the star formation in NGC~1367, our result of no radio AGNs in passive spiral galaxies suggests that radio AGNs are no longer present in spiral galaxies after their star formation is quenched. If radio mode feedback continues to regulate star formation in passive early-type galaxies, this is not the case in passive spiral galaxies.

%% PARA: NGC 1371 impacts
NGC~1367 is not the first example of low-SFR spiral galaxies with radio emission primarily powered by an AGN. \citet{kav2015} identified AGNs with radio luminosities of $10^{21} - 10^{24} \mathrm{~W/Hz}$ in spiral host galaxies with SFRs of $0.14 - 25 ~M_\odot \rm{~yr^{-1}}$, which are higher in both radio luminosity and SFR than NGC~1367. While WISE infrared colours filter out most star-forming galaxies, the example of NGC~1367 highlights the importance of removing star-forming contaminants because erroneously classifying NGC~1367 as a passive galaxy would significantly alter our conclusions. 

\end{document}